\documentclass[%
reprint,
superscriptaddress,
showpacs,
 amsmath,amssymb,
aps,
prl,
floatfix,
]{revtex4-1}

\usepackage{gensymb}
\usepackage{graphicx}
\usepackage{dcolumn}
\usepackage{bm}
\usepackage{hyperref}
\usepackage[mathlines]{lineno}

\usepackage{tikz}
\usetikzlibrary{positioning}
\usetikzlibrary{calc}

\hypersetup{
    pdfnewwindow=true,      
    colorlinks=true,       
    linkcolor=blue,          
    citecolor=blue,        
    filecolor=blue,      
    urlcolor=blue           
}

\newcommand{\TT}{\textrm{TT}}
\newcommand{\dptt}{\delta p_\TT}
\newcommand{\nm}{\nu_{\mu}}
\newcommand{\nee}{\nu_\textrm{e}}
\newcommand{\anee}{\bar{\nu}_\textrm{e}}
\newcommand{\anm}{\bar{\nu}_{\mu}}
\newcommand{\enu}{E_{\nm}}
\newcommand{\ztt}{\vec{z}_\TT}
\newcommand{\res}{\Delta^{++}}
\newcommand{\difd}{\textrm{d}}
\newcommand{\dEdx}{\difd E/\difd x}
\newcommand{\proton}{\textrm{p}}
\newcommand{\neutron}{\textrm{n}}
\newcommand{\hydrogen}{\textrm{H}}

\newcommand{\lepton}{l}
\newcommand{\leptonminus}{\lepton^{-}}
\newcommand{\leptonplus}{\lepton^{+}}
\newcommand{\neutrino}{\nu}
\newcommand{\antineutrino}{\bar{\nu}}
\newcommand{\nna}{\neutrino/\antineutrino}
\newcommand{\lpm}{\lepton^{\mp}}
\newcommand{\pf}{p_\proton}

\newcommand{\ox}{\affiliation{Department of Physics, Oxford University, Oxford, United Kingdom}}
\newcommand{\ral}{\affiliation{STFC, Rutherford Appleton Laboratory, Harwell Oxford, United Kingdom}}

\begin{document}

\preprint{XXX}

\title{
                            Reconstruction of Energy Spectra of Neutrino Beams Independent of Nuclear Effects
}

\author{X.-G.~Lu}\email{Xianguo.Lu@physics.ox.ac.uk}\ox
\author{D.~Coplowe}\ox
\author{R.~Shah}\ox\ral
\author{G.~Barr}\ox
\author{D.~Wark}\ox\ral
\author{A.~Weber}\ox\ral

\date{\today}

\begin{abstract}

 We propose a new technique which enables an event-by-event selection of neutrino-hydrogen  interactions in multi-nuclear targets and thereby  allows application of hydrogen as targets  in  experiments with neutrino beams without involving cryogenics or high pressure  hydrogen gas. This technique could significantly improve the reconstruction of the neutrino energy spectra.   
  Since it allows a separation between hydrogen and the accompanying nuclei, this technique also enables us to measure  nuclear effects in neutrino interactions directly.

\end{abstract}


\maketitle


The ability to measure the  energy spectrum  of a neutrino~\cite{footnote:nuantinu} beam has many physics implications~\cite{Adams:2013qkq, Abe:2015zbg}. The accuracy of the  measurement depends on the spectral shape, the energy reconstruction  and  the understanding of the cross sections of the processes by which the neutrino interactions are detected. The conventional 
 measurement of the energy spectra of neutrino beams  is via  charged-current quasi-elastic  scattering (CCQE) on nucleons: $\neutrino+\neutron\to\leptonminus+\proton$ and $\antineutrino+\proton\to\leptonplus+\neutron$, where $\nna$, $\neutron$, $\proton$ and $\lpm$ stand for neutrino/anti-neutrino, neutron, proton and the corresponding charged leptons, respectively. The neutrino energy can be  calculated  using the lepton momentum, assuming a static  nucleon in the initial state~\cite{Abe:2013hdq, Abe:2015zbg}. With a nuclear target, the accuracy is limited by the binding energy and the Fermi motion (FM) of the  nucleon, both subject to large  fluctuations, and by other initial-state uncertainties. When the momentum of the final-state nucleon is measured, the neutrino energy can be reconstructed by summing all the final-state particle momenta. However, the kinematics of the final-state nucleon are altered by final-state interactions (FSIs)  as the  nucleon re-interacts with the cold nuclear medium before leaving the target nucleus. FSIs can be so strong that the  nucleus is excited or even breaks up, emitting low momentum particles such as nucleons,  photons and  pions, which are  stopped near the vertex and not detected in tracking detectors,  leading to greater bias in the reconstructed energy. Therefore a third approach (not restricted to CCQE) is to sum the lepton energy and the visible energy of the hadronic system~\cite{Ayres:2004js, Michael:2006rx, Adams:2013qkq},  which is limited by a reduced influence from nucleon initial-state uncertainties and  by the systematics in measuring the energy of neutral particles. Among those initial-state uncertainties,  multi-nucleon correlations~\cite{Egiyan:2005hs, Shneor:2007tu, AguilarArevalo:2010zc, AguilarArevalo:2013hm, Martini:2010ex, Nieves:2011yp} are under intense study. Such nuclear modifications make the calculation of the  CCQE cross section difficult.  Furthermore, in non-CCQE interactions such as resonance production, the final-state pions can re-scatter, exchange charge, or be  absorbed in  the nuclear medium~\cite{Ashery:1986nt}. Such background events are often mis-identified as CCQE  due to the identical final-state particles and therefore introduce an ambiguity in the cross section definition. More details about   neutrino-nucleus interactions  can be found in Refs.~\cite{Gallagher:2011zza, Formaggio:2013kya} and references therein.

Hydrogen is the ideal target for reconstructing the neutrino energy because of the absence of these nuclear effects, however a hydrogen target with high mass is technically impracticable.  In this work, we propose a solution which is using
  a spatial symmetry  in the final-state kinematics in charged-current (CC) resonance production to isolate hydrogen events in targets with a mixture of nuclei.  
This would allow the reconstruction of the beam energy spectrum only limited by the knowledge of the cross section on hydrogen, which is much better understood than those on nuclei.

\begin{figure}
\begin{center}

\begin{tikzpicture}[scale=0.5]

\coordinate (O) at (1,0,0);
\draw[very thin,-latex] (O) -- +(2, 0,  0) node [below] {$x$};
\draw[very thin,-latex] (O) -- +(0,  2, 0) node [left] {$y$};
\draw[very thin,-latex] (O) -- +(0,  0, 2) node [left] {$z$};

\coordinate (NDP) at (0,0,-6);
\coordinate (VERT) at (6,0,-6);

\draw[thick] (NDP) --  ($ (VERT) + (-1,0,0) $);
\draw[thick,opacity=0.6,-latex]  ($ (VERT) + (-1,0,0) $) -- (VERT) node [above left = 0.0cm and 2cm,opacity=1] {$\vec{p}_{\nu/\bar{\nu}}$};
\draw[thick,opacity=0.6] (VERT) -- ($ (VERT) + (2.4,4,0) $);
\draw[thick,-latex] ($ (VERT) + (2.4,4,0) $) -- ($ (VERT) + (3,5,0) $) node [left = .1cm] {$\vec{p}_{l^\mp}$};
\draw[thick,color=blue][-latex] (VERT) -- ($ (VERT) + (1,-3, 3) $) node [below right = 0cm and 0.0cm] {$\vec{p}_\textrm{X}$};
\draw[thick,color=blue,opacity=0.4][-latex] (VERT) -- ($ (VERT) + (2,-2,-3) $) node [below right = 0cm and 0.0cm] {$\vec{p}_\textrm{Y}$};

\draw node [above right = 0.7 cm and 4.2cm] {p};
\draw[very thin,-latex] (VERT) -- ($ (VERT) + (0,0, 7.5) $) node [below  = 0.cm ] {$\vec{z}_\textrm{TT}$};
\draw[very thin, opacity=0.4] (VERT) -- ($ (VERT) + (0,0, -7.5) $);

\draw[thick,color=red][-latex] (VERT)  -- ($ (VERT) + (0,0, 3) $) node [ above left = 0.02cm and 0.cm] {$p^\textrm{X}_\textrm{TT}$};
\draw[thick,color=red,opacity=0.4][-latex] (VERT)  -- ($ (VERT) + (0,0, -3) $) node [ right = 0.2cm] {$p^\textrm{Y}_\textrm{TT}$};

\draw[help lines,style=dashed]  ($ (VERT) + (1,-3, 3) $) -- ($ (VERT) + (0,0, 3) $);
\draw[help lines,style=dashed,opacity=0.4]  ($ (VERT) + (2,-2,-3) $)  --  ($ (VERT) + (0,0, -3) $);

\draw[fill=black!90!white,opacity=0.05] ($ (VERT) + (5,4,0) $) -- ($ (VERT) + (5,-2,0) $) -- ($ (VERT) + (-1,-2,0) $) -- ($ (VERT) + (-1,4,0) $);

\draw node [above right = 3cm and -1.7cm] {\{X, Y\}};
\draw node [above right = 2.5cm and -1.7cm] {= \{p, $\pi^+$\} for $\nu+\textrm{p}\rightarrow l^-+\Delta^{++}$};
\draw node [above right = 2cm and -1.7cm] {or \{p, $\pi^-$\} for $\bar{\nu}+\textrm{p}\rightarrow l^++\Delta^0$  };

\end{tikzpicture}

\caption{Schematic illustration of the double-transverse kinematics. The incoming and outgoing particle momenta are represented by $\vec{p}_{\nu}$ and $\vec{p}_{l}$ ,~$\vec{p}_\textrm{p}$ and~$\vec{p}_{\pi}$, respectively. The double-transverse momentum imbalance, $\dptt$ is given by  $p_\TT^\proton+p_\TT^\pi$ with respect to the axis $\ztt$ defined by $\vec{p}_{\neutrino}\times\vec{p}_\lepton$.    }\label{fig:dpttdef}
\end{center}
\end{figure}
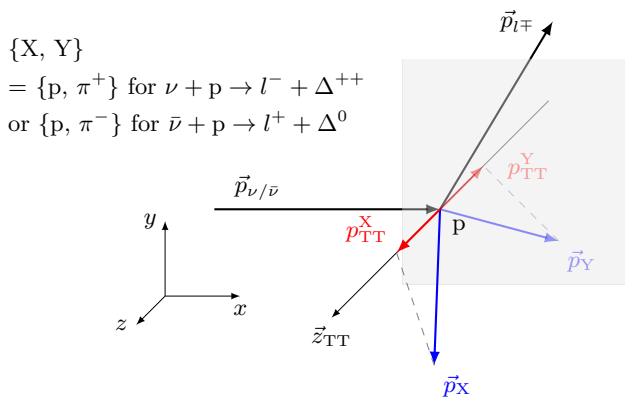

Delta resonances $\Delta(1232)$ can be produced in CC interactions on hydrogen when the neutrino energy is above threshold (about 0.34~GeV and 0.49~GeV for $\nee$ and $\nm$, respectively). Consider a $\nee$ or $\nm$  interaction on a proton $\neutrino+\proton\to\leptonminus+\res$, where the $\res$ decays to a proton and a positive pion, $\pi^{+}$. We define a double-transverse axis $\ztt\equiv\vec{p}_{\neutrino}\times\vec{p}_\lepton/\left|\vec{p}_{\neutrino}\times\vec{p}_\lepton\right|$, which is by construction perpendicular to both the neutrino and charged lepton momenta, $\vec{p}_{\neutrino}$ and $\vec{p}_\lepton$. On projecting the proton and pion momenta, $\vec{p}_\proton$ and $\vec{p}_\pi$,  onto $\ztt$, $p_\TT^\proton\equiv\vec{p}_\proton\cdot\ztt$,  $p_\TT^{\pi}\equiv\vec{p}_\pi\cdot\ztt$, one has the double-transverse momentum imbalance $\dptt\equiv p_\TT^\proton+p_\TT^\pi$ (see Fig.~\ref{fig:dpttdef} for a schematic illustration).  In the absence of nuclear effects,  as is expected for a hydrogen target, $\dptt$ is zero, whereas $\dptt\neq0$  in the presence of FM and FSI in a nuclear target. This is independent of the neutrino energy and the resonance kinematics. The  $\dptt$  for a nuclear target has the following properties:
(1) It is distributed symmetrically around zero because the initial proton motion  and the decay kinematics of the  resonance are uncorrelated to  $\ztt$ (except for  uncommon cases such as  polarized spatially-asymmetric nuclear targets, or if the   detection acceptance varies  for different final-state particles).
(2) Since FM is isotropic, for a given initial proton momentum $\pf$ (up to about 200~MeV$/c$ for carbon~\cite{Povh:1995mua}, for example),  $\dptt$ is broadened from 0 to the same order of magnitude as $\pf$. The randomness of FM further smears out $\dptt$.
(3) The resonance and the decay products experience FSI. Such modification of the kinematics  further adds to the broadening of the $\dptt$ distribution.


\begin{figure}
\begin{center}
\includegraphics[width=\columnwidth]{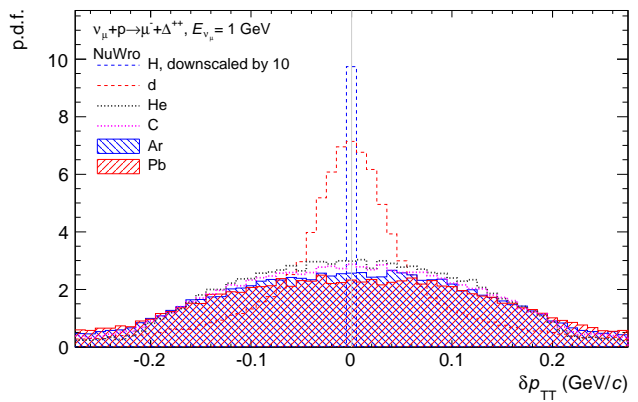}
\caption{Probability density function (p.d.f.) of the  double-transverse kinematic imbalance $\dptt$  generated by NuWro~\cite{Golan:2012wx}   for hydrogen, deuteron, helium, carbon, argon and lead targets with  neutrino energy 1~GeV. The width of the hydrogen distribution is due to the finite bin width. }\label{fig:neutdptt}
\end{center}
\end{figure}

The difference between the $\dptt$ shapes for  hydrogen and nuclear targets is dramatic (see Fig.~\ref{fig:neutdptt}). As can be seen in the figure, the shapes of the nuclear distributions predicted by NuWro~\cite{Golan:2012wx} vary only slightly among nuclei heavier than deuteron.  Therefore for a  multi-nuclear target  with hydrogen, assuming perfect detector response,  one  expects a hydrogen signal at $\dptt=0$ on top of a  symmetric nuclear background that is about 200~MeV$/c$ wide.  At the reconstruction level, the shape of the hydrogen peak,  which is still symmetric, is solely determined by the  detector response. The nuclear  background contamination  under the hydrogen peak depends on the background shape and  the resonance production cross section ratio between the nucleus and hydrogen, which equals roughly the  atomic number of the nucleus modulo nuclear effects. Improving the detector resolution~\cite{flexi} will lead to  a strong signal enhancement and eventually an event-by-event selection of hydrogen interactions.  Once the hydrogen interactions are selected, the neutrino energy can be reconstructed by summing the final-state energy~\cite{longi}. The energy reconstruction quality is solely determined by  detector response  and not limited by nuclear effects.

The advantage of $\dptt$ is clear when compared to other characteristic variables in the  interaction such as the invariant mass of $\Delta(1232)$ and the total transverse momentum. The former  has an  irreducible Breit-Wigner width of about 117~MeV$/c^2$~\cite{Agashe:2014kda} and therefore has no sensitivity to reject nuclear background; the latter, which is intrinsically also zero for a hydrogen target, is asymmetric after reconstruction and in general has a long tail resembling the Landau distribution~\cite{landau} due to detector  effects.

 For the anti-neutrino interaction on a proton $\antineutrino+\proton\to\leptonplus+\Delta^0$, where $\Delta^0$ decays to $\proton+\pi^{-}$, the previous definition and discussions directly apply. This  similarity  enables  highly consistent measurements of  neutrino and anti-neutrino energy spectra.

Given limited detector resolution, it is important to minimize background that has non-zero $\dptt$. Because FSI can lead to soft nuclear emission, measuring the vertex energy~\cite{Fields:2013zhk, Fiorentini:2013ezn} allows tagging and rejecting interactions on other nuclei. And since  pion FSI can modify the final states,  the hydrogen signal purity could be enhanced by choosing target materials whose nuclear part has a large pion FSI cross section. Non-exclusive background such as multiple pion production can be rejected by vetoing electromagnetic processes and  neutral particles~\cite{Allan:2013ofa}.

For existing experiments, given a well understood  detector response, it may be feasible to perform a combined fit  to the center  region of the $\dptt$ distribution for a mixed target, where the  hydrogen shape is fixed and the  background modeling follows the general properties described above. If the signal width is at the sub-hundred MeV level, the fit may not be sensitive to the detail of the complicated nuclear tails. The  cross section of the  resonance production on hydrogen, which is independent of nuclear effects, can be obtained from the signal part. The yield ratio between the hydrogen signal and the remaining contribution from the other target nuclei  is a precise measurement of the associated nuclear effects with  cancellation of detection acceptance and efficiencies for both targets. Such measurement of the resonance production on hydrogen and nuclei  should largely improve the understanding of the production mechanism and its modification by the nuclear medium.

More challenging background processes are those with intrinsic zero $\dptt$. One  type of such background is the exclusive processes which have identical final states as the signal, such as higher mass resonances and non-resonant production~\cite{cohbk}. Those processes do not affect the neutrino energy reconstruction, but make it difficult to define the interaction cross section which is needed to determine the beam energy spectrum. To distinguish among underlying processes, the detailed  interaction kinematics could be used,  such as the invariant mass of the hadronic system $W$ and  the squared four-momentum exchange to the proton $t$. This is feasible thanks to the essential features --- being exclusive and nuclear effect-independent ---  that make the neutrino energy reconstruction precise. Because the cross sections have different dependence on $W$ and $t$, efficient separation should be possible.  Since the relevant final state kinematics do not depend on the identity of the intermediate state, alternatively one could extend the definition of the production channel to include all contributions that have exclusive $\proton\pi^+$ final states and calculate the corresponding cross section~\cite{Adler:1968tw, Rein:1980wg}. 
 Another type of background is the $\nm$ ($\anm$) contamination in the $\anm$ ($\nm$) CC interaction. Such ``wrong sign'' background occurs more often in an anti-neutrino beam created in a proton accelerator. Because of the similar particle identification (PID) signals for the muon and pion, the $\mu^{-}+\pi^{+}+\proton$ final states from the $\nm$ background might be misidentified  in the event selection as $\pi^{-}+\mu^{+}+\proton$ from $\anm$, and  vice versa. For a large enough detection volume, the properties of stopped pions can be used to enhance the pion identification. In addition to measuring  trajectory ``kinks"~\cite{Abelev:2014ffa} and the Michel electrons~\cite{AguilarArevalo:2011sz},  negative pions that are  stopped and absorbed by nuclei give rise to soft nuclear emission which can be measured by calorimetry~\cite{Marin:1998zf}.  The lepton, $\pi^{+}$ and $\pi^{-}$  signatures are different and   serve as an important tool not only for $\lepton/\pi$ separation, but also for rejecting ``wrong sign'' contamination. In addition,  variables like $W$, which is calculated from the true $\proton\mu$ system due to wrong PID,    can provide background rejection power due to the unphysical kinematic combination~\cite{Campbell:1973wg}.

In this work,  MC simulation of  the T2K  ND280 detector~\cite{Abe:2011ks}  is used to demonstrate the measurement of CC resonance production in neutrino-hydrogen interactions, $\nm+\hydrogen\to\mu^{-}+\res$.  Neutrino interactions in  ND280 are simulated using the Neut event generator~\cite{Hayato:2002sd}.  The event reconstruction uses the first  fine-grained detector (FGD~\cite{Amaudruz:2012esa}) as a plastic scintillator (polystyrene) interaction target and the neighboring gaseous time projection chamber  (TPC~\cite{Abgrall:2010hi})   downstream   to measure the momenta and specific energy loss ($\dEdx$) of the final-state particles. 
 There are  about 300 true $\res$ events on hydrogen in the neutrino energy range $0.5<\enu<5$~GeV in the reconstructed sample.

\begin{figure}
\begin{center}
\includegraphics[width=\columnwidth]{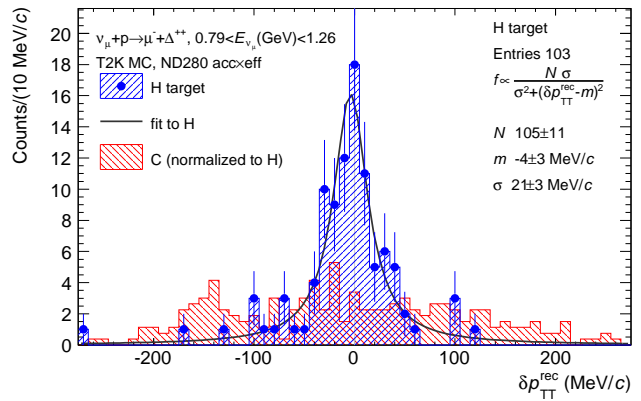}
\caption{Distributions of the reconstructed $\dptt$  in the ND280 acceptance (detection efficiencies apply) simulated for hydrogen and carbon  target nuclei in the neutrino energy range \mbox{0.79 -- 1.26~GeV}. Vertical bars are MC statistical errors. The hydrogen signal is fit to a Cauchy distribution. The carbon distribution is area-normalized to the hydrogen.}\label{fig:dpttrec}
\end{center}
\end{figure}

\begin{figure}
\begin{center}
\includegraphics[width=\columnwidth]{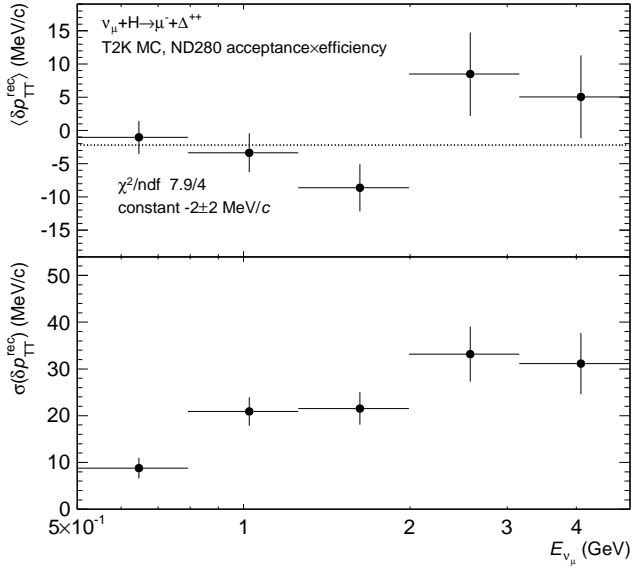}
\caption{The Cauchy mean (upper) and width (lower) of the reconstructed $\dptt$ as a function of the neutrino energy. Vertical bars are MC statistical errors,  while horizontal error bars stand for the bin span of the neutrino energy. A fit of a constant is applied to the mean values.}\label{fig:dpttenu}
\end{center}
\end{figure}

The  distribution of the reconstructed double-transverse kinematic imbalance $\dptt$ in the ND280 acceptance (detection efficiencies apply)  is shown in Fig.~\ref{fig:dpttrec}. Due to the absence of nuclear effects, for hydrogen the resolution is determined by the detector response,  which can be  described in the simulation by a Cauchy function $1/N\cdot\difd N/\difd\dptt=1/\pi\cdot\sigma/[\sigma^2+(\dptt-m)^2]$ with the event count $N$, mean $m$ and width $\sigma$. The  resolution is shown to be significantly smaller than the nuclear broadening in the carbon target nuclei. The nuclear rejection factor defined as the efficiency ratio between hydrogen and carbon 
 is about 3.7 (2.3) in the 1 (3) $\sigma$  interval.  
 The reconstruction performance of $\dptt$ as a function of the neutrino energy is shown in Fig.~\ref{fig:dpttenu}. It indicates a better hydrogen selection  at lower energy.

\begin{figure}
\begin{center}
\includegraphics[width=\columnwidth]{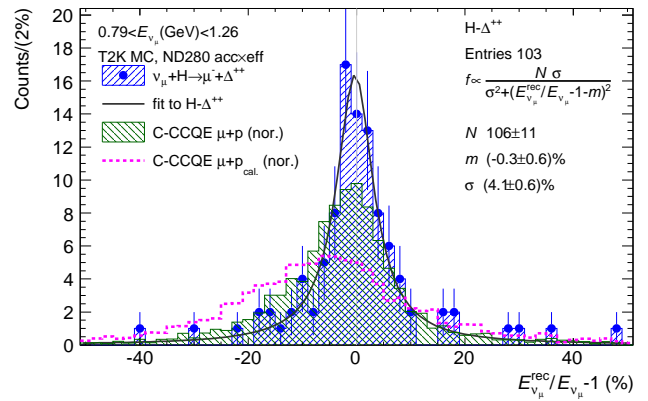}
\caption{Deviation of the reconstructed neutrino energy from the true value. Reconstruction with the CC resonance production on hydrogen targets is compared to the ones using  CCQE on carbon: one summing both muon and proton reconstructed kinematics (``$\mu+\proton$"), the other using the muon  to calculate the proton momentum assuming a static initial neutron (``$\mu+\proton_\textrm{cal.}$")~\cite{Abe:2013hdq}. The carbon distributions are area-normalized to the hydrogen.  }\label{fig:enurec}
\end{center}
\end{figure}

\begin{figure}
\begin{center}
\includegraphics[width=\columnwidth]{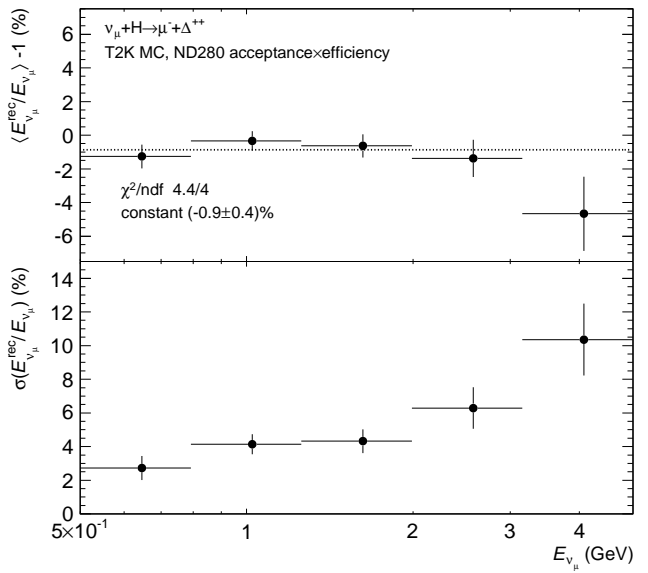}
\caption{Neutrino energy scale (upper) and reconstruction resolution (lower) via CC resonance production on hydrogen.}\label{fig:enuscale}
\end{center}
\end{figure}

From the MC true hydrogen-resonance events  in the reconstructed sample, the neutrino energy $\enu$ is calculated directly using the kinematics of all final-state particles. The uncertainty in an example $\enu$ bin is shown in Fig.~\ref{fig:enurec}.   Like $\dptt$,  it is determined by pure detector resolution  due to the  free and static proton target and  can  be described by a Cauchy function. As a comparison, the neutrino energy in the same range is reconstructed with all final-state kinematics from CCQE interactions on carbon assuming static neutron targets. This simple  reconstruction suffers from nuclear effects in addition to the detector resolution and is biased by about  $-2\%$ with a large spread. In addition, the neutrino energy distribution reconstructed only with the muon kinematics~\cite{Abe:2013hdq, ccqeformu} shows worse performance. The comparison between both CCQE methods   indicates that, for the relatively weak FSI predicted in the current model, an independent measurement of the proton kinematics help improve the $\enu$ resolution even in the presence of FSI.   The simulated $\enu$ detector response (the Cauchy mean and width) is further shown as a function of the true energy in Fig.~\ref{fig:enuscale}.
 The energy scale is seen to be constant with a bias of about -1\%, and the resolution  slowly increases  from about 3\% to 10\%  in  the \mbox{0.5 -- 5~GeV} region. The worsening of resolution at high energy is general for a tracker measurement, in contrast to calorimetry~\cite{Ayres:2004js, Michael:2006rx, Adams:2013qkq}.

Given the fact that  ND280  was optimized to measure CCQE interactions and designed to achieve a performance that was to be limited by the nucleon initial-state uncertainties in the nuclear target~\cite{Amaudruz:2012esa, Abgrall:2010hi}, alternative optimization and state-of-the-art technology may allow  better performance. Since the CC resonance  production cross section  rises to  a maximum  at a neutrino energy greater than about 3~GeV~\cite{Agashe:2014kda}, the high energy neutrinos produced in the NuMI~\cite{Anderson:1998zza} and LBNF~\cite{Adams:2013qkq} beam lines are optimal to realize the proposed method.  
 It would be  challenging, yet very attractive, to combine the proposed use of  hydrogen targets and the liquid argon TPC projects~\cite{Chen:2007ae, Anderson:2012vc, Adams:2013uaa, Adams:2013qkq, Antonello:2013ypa} which are to realize  superb tracking and calorimetry performance on a massive scale. 
 Finally, it would be interesting to demonstrate the method with anti-neutrino beams as well as for $\nee/\anee$  interactions in oscillation appearance and investigate the impact on the physics programs in  future experiments,  considering also the advantage of the identical target, interaction kinematics and phase space in the proposed channels $\nna+\proton\rightarrow\lpm+\Delta\rightarrow\lpm+\proton+\pi^\pm$, in comparison to the conventional CCQE interactions.

The authors would like to express their gratitude to the T2K experiment for the use of the full MC production and the excellent analysis software in this study, and to  T.~Dealtry, K.~Duffy, R.~Guenette, A.~Jacob, K.~McFarland, V.~Paolone, L.~Pickering, F.~Sanchez, H.~Tanaka, Y.~Uchida, M.~Wascko, C.~Wilkinson, M.~Yokoyama and M.~Zito for  helpful discussions.  We are grateful to R.~Guenette and Y.~Uchida for useful suggestions on improving the manuscript. This work is supported by the UK Science and Technology Facilities Council.


\begin{thebibliography}{00}

\bibitem{footnote:nuantinu}
In this paper, unless otherwise specified, \emph{neutrino} refers to both \emph{neutrino} and \emph{anti-neutrino}.

\bibitem{Adams:2013qkq} 
  C.~Adams {\it et al.}  [LBNE Collaboration],
  BNL-101354-2013-JA, BNL-101354-2014-JA, FERMILAB-PUB-14-022, LA-UR-14-20881 (2013).


\bibitem{Abe:2015zbg} 
  K.~Abe {\it et al.}  [Hyper-Kamiokande Proto- Collaboration],
  PTEP {\bf 2015}, no. 5, 053C02 (2015).


\bibitem{Abe:2013hdq} 
  K.~Abe {\it et al.}  [T2K Collaboration],
  Phys.\ Rev.\ Lett.\  {\bf 112}, 061802 (2014).


\bibitem{Ayres:2004js} 
  D.~S.~Ayres {\it et al.}  [NOvA Collaboration],
  hep-ex/0503053.

\bibitem{Michael:2006rx} 
  D.~G.~Michael {\it et al.}  [MINOS Collaboration],
  Phys.\ Rev.\ Lett.\  {\bf 97}, 191801 (2006).

\bibitem{Egiyan:2005hs}
  K.~S.~Egiyan {\it et al.}  [CLAS Collaboration],
  Phys.\ Rev.\ Lett.\  {\bf 96} (2006) 082501.

\bibitem{Shneor:2007tu} 
  R.~Shneor {\it et al.}  [Jefferson Lab Hall A Collaboration],
  Phys.\ Rev.\ Lett.\  {\bf 99}, 072501 (2007).

\bibitem{AguilarArevalo:2010zc} 
  A.~A.~Aguilar-Arevalo {\it et al.}  [MiniBooNE Collaboration],
  Phys.\ Rev.\ D {\bf 81}, 092005 (2010).


\bibitem{AguilarArevalo:2013hm} 
  A.~A.~Aguilar-Arevalo {\it et al.}  [MiniBooNE Collaboration],
  Phys.\ Rev.\ D {\bf 88}, no. 3, 032001 (2013).

\bibitem{Martini:2010ex} 
  M.~Martini, M.~Ericson, G.~Chanfray and J.~Marteau,
  Phys.\ Rev.\ C {\bf 81}, 045502 (2010).

\bibitem{Nieves:2011yp} 
  J.~Nieves, I.~Ruiz Simo and M.~J.~Vicente Vacas,
  Phys.\ Lett.\ B {\bf 707}, 72 (2012).


\bibitem{Ashery:1986nt} 
  D.~Ashery and J.~P.~Schiffer,
  Ann.\ Rev.\ Nucl.\ Part.\ Sci.\  {\bf 36}, 207 (1986).


\bibitem{Gallagher:2011zza}
  H.~Gallagher, G.~Garvey and G.~P.~Zeller,
  Ann.\ Rev.\ Nucl.\ Part.\ Sci.\  {\bf 61} (2011) 355.


\bibitem{Formaggio:2013kya} 
  J.~A.~Formaggio and G.~P.~Zeller,
  Rev.\ Mod.\ Phys.\  {\bf 84}, 1307 (2012).


\bibitem{Povh:1995mua} 
  B.~Povh, K.~Rith, C.~Scholz and F.~Zersche,
  ``Particles and nuclei: An Introduction to the physical concepts,''
  Berlin, Germany: Springer (2009) 428 p.

\bibitem{Golan:2012wx} 
  T.~Golan, C.~Juszczak and J.~T.~Sobczyk,
  Phys.\ Rev.\ C {\bf 86}, 015505 (2012).


\bibitem{flexi}
Concerning the resolution, there is  some flexibility to optimize the signal selection: any of the final-state momentum directions can be used  to construct $\ztt$ and only the momenta of the other two particles are required for reconstructing $\dptt$.

\bibitem{longi}
Depending on the tracking, calorimetry and particle identification performance of a detector, alternative calculation of the  neutrino energy by summing the longitudinal components of the final-state momenta could be considered.

\bibitem{Agashe:2014kda} 
  K.~A.~Olive {\it et al.}  [Particle Data Group Collaboration],
  Chin.\ Phys.\ C {\bf 38}, 090001 (2014).

\bibitem{landau}
L.~D.~Landau, ``On the energy loss of fast particles by ionization," J. Phys. 
(USSR) 8 (1944) 201. See also D.~Ter~Haar (ed.), ``Collected Papers of L.~D.~Landau," 
Gordon \& Breach Science (1965) p. 417.


\bibitem{Fields:2013zhk} 
  L.~Fields {\it et al.}  [MINERvA Collaboration],
  Phys.\ Rev.\ Lett.\  {\bf 111}, no. 2, 022501 (2013).

\bibitem{Fiorentini:2013ezn} 
  G.~A.~Fiorentini {\it et al.}  [MINERvA Collaboration],
  Phys.\ Rev.\ Lett.\  {\bf 111}, 022502 (2013).

\bibitem{Allan:2013ofa} 
  D.~Allan {\it et al.}  [T2K UK Collaboration],
  JINST {\bf 8}, P10019 (2013).

\bibitem{cohbk}
In coherent pion production on hydrogen, the  slowly recoiling proton usually escapes detection. Therefore  coherent pion production  in general is not a background process.


\bibitem{Adler:1968tw} 
  S.~L.~Adler,
  Annals Phys.\  {\bf 50}, 189 (1968).

\bibitem{Rein:1980wg} 
  D.~Rein and L.~M.~Sehgal,
  Annals Phys.\  {\bf 133}, 79 (1981).

\bibitem{Abelev:2014ffa} 
  B.~B.~Abelev {\it et al.}  [ALICE Collaboration],
  Int.\ J.\ Mod.\ Phys.\ A {\bf 29}, 1430044 (2014).

\bibitem{AguilarArevalo:2011sz} 
  A.~A.~Aguilar-Arevalo {\it et al.}  [MiniBooNE Collaboration],
  Phys.\ Rev.\ D {\bf 84}, 072005 (2011).


\bibitem{Marin:1998zf} 
  A.~Marin, J.~Diaz, R.~Averbeck, A.~Doppenschmidt, S.~Hlavac, R.~Holzmann, F.~Lefevre and A.~Schubert {\it et al.},
  Nucl.\ Instrum.\ Meth.\ A {\bf 417}, 137 (1998).


\bibitem{Campbell:1973wg} 
  J.~Campbell, G.~Charlton, Y.~Cho, M.~Derrick, R.~Engelmann, J.~Fetkovich, L.~Hyman and K.~Jaeger {\it et al.},
  Phys.\ Rev.\ Lett.\  {\bf 30}, 335 (1973).

\bibitem{Abe:2011ks} 
  K.~Abe {\it et al.}  [T2K Collaboration],
  Nucl.\ Instrum.\ Meth.\ A {\bf 659}, 106 (2011).

\bibitem{Hayato:2002sd} 
  Y.~Hayato,
  Nucl.\ Phys.\ Proc.\ Suppl.\  {\bf 112}, 171 (2002).


\bibitem{Amaudruz:2012esa} 
  P.~A.~Amaudruz {\it et al.}  [T2K ND280 FGD Collaboration],
  Nucl.\ Instrum.\ Meth.\ A {\bf 696}, 1 (2012).

\bibitem{Abgrall:2010hi} 
  N.~Abgrall {\it et al.}  [T2K ND280 TPC Collaboration],
  Nucl.\ Instrum.\ Meth.\ A {\bf 637}, 25 (2011).

\bibitem{ccqeformu}
In water Cherenkov detectors, the final-state protons are mostly below detection threshold. In such cases the neutrino energy can only be estimated using the muon kinematics.  

\bibitem{Anderson:1998zza} 
  K.~Anderson, B.~Bernstein, D.~Boehnlein, K.~R.~Bourkland, S.~Childress, N.~Grossman, J.~Hylen and C.~James {\it et al.},
  FERMILAB-DESIGN-1998-01.

\bibitem{Chen:2007ae} 
  H.~Chen {\it et al.}  [MicroBooNE Collaboration],
  FERMILAB-PROPOSAL-0974.

\bibitem{Anderson:2012vc} 
  C.~Anderson, M.~Antonello, B.~Baller, T.~Bolton, C.~Bromberg, F.~Cavanna, E.~Church and D.~Edmunds {\it et al.},
  JINST {\bf 7}, P10019 (2012).

\bibitem{Adams:2013uaa} 
  C.~Adams {\it et al.}  [LArTPC Collaboration],
  arXiv:1309.7987 [physics.ins-det].

\bibitem{Antonello:2013ypa} 
  M.~Antonello, B.~Baibussinov, V.~Bellini, H.~Bilokon, F.~Boffelli, M.~Bonesini, E.~Calligarich and S.~Centro {\it et al.},
  arXiv:1312.7252 [physics.ins-det].

\end{thebibliography}
\end{document}